\documentclass[twocolumn,nopacs,amsmath,amssymb,superscriptaddress,prl]{revtex4-1}

\usepackage[utf8]{inputenc}
\usepackage{amsmath}
\usepackage{amsfonts}
\usepackage{amssymb}
\usepackage{graphicx} 
\usepackage{dcolumn}
\usepackage{bm}
\usepackage{natbib}
\usepackage{braket}
\usepackage{xcolor}

\newcommand{\1}{\uparrow} 
\newcommand{\2}{\downarrow} 
\newcommand{\be}{\begin{equation}}
\newcommand{\ee}{\end{equation}}

\begin{document}

\title{Tunable three-body interactions in driven two-component Bose-Einstein condensates}

\author{A. Hammond}
\affiliation{Laboratoire Charles Fabry, UMR 8501, Institut d'Optique, CNRS, Universit\'e Paris-Saclay, Avenue Augustin Fresnel, 91127 Palaiseau CEDEX, France}
\author{L. Lavoine}
\affiliation{Laboratoire Charles Fabry, UMR 8501, Institut d'Optique, CNRS, Universit\'e Paris-Saclay, Avenue Augustin Fresnel, 91127 Palaiseau CEDEX, France}
\author{T. Bourdel}
\email[Corresponding author: ]{thomas.bourdel@institutoptique.fr}
\affiliation{Laboratoire Charles Fabry, UMR 8501, Institut d'Optique, CNRS, Universit\'e Paris-Saclay, Avenue Augustin Fresnel, 91127 Palaiseau CEDEX, France}

\date{\today}

\begin{abstract}
We propose and demonstrate the appearance of an effective attractive three-body interaction in coherently-driven two-component Bose Einstein condensates. It originates from the spinor degree of freedom that is affected by a two-body  mean-field shift of the driven transition frequency. Importantly, its strength can be controlled with the Rabi-coupling strength and it does not come with additional losses. In the experiment, the three-body interactions are adjusted to play a predominant role in the equation of state of a cigar-shaped trapped condensate. This is confirmed though two striking observations: a downshift of the radial breathing mode frequency and the radial collapses for positive values of the dressed-state scattering length.
\end{abstract}

\maketitle

Thanks to their extreme diluteness, particles in ultracold gases dominantly interact pairwise. At low temperatures, the thermal de Broglie wavelength is larger than the range of the Van der Waals potential $R_e$, and the two-body interaction can accurately be replaced by a contact potential \cite{Pitaevskii03}. Morevover, the only parameter characterizing the interaction, {\it i.e.} the scattering length $a$ can be tuned via scattering resonances \cite{Chin10}. Thanks to these properties, ultracold gases are ideal candidates to quantitatively explore quantum many-body physics with pairwise interactions \cite{Bloch08}. For example, the superfluid to Mott transition \cite{Greiner02} or the BEC-BCS crossover \cite{Regal04, Zwierlein04, Bartenstein04, Bourdel04, Zwerger16} have been studied. 

Although three-body interactions are usually a small correction as compared to two-body interactions in dilute gases, their consideration has a long history \cite{Wu59, Kohler02, Hammer13}. Theoretically, they lead to interesting non-linear dynamics \cite{Gammal00, Gammal00b, Kumar10, Crosta12, Al-Jibbouri13, Killip17, Zloshchastiev17} and to the appearance of droplets \cite{Bulgac02}. At low temperatures, a three-body interaction is characterized by a scattering hypervolume $D$ \cite{Tan08}. $D$ is a complex number whose real part is associated with an energy shift and its imaginary part with three-body losses. Enhancement of three-body interactions, i.e. of the real part of $D$ is expected close to resonances due to energy coincidence with weakly bound three-body states \cite{Bulgac02, Tan08, Zhu17, Incao18, Mestrom19, Zwerger19}. Unfortunately, typical interatomic interaction potentials possess numerous  deeply bound two-body states and the enhancement of the real part of $D$ comes together with a concomitant increase of its imaginary part due to three-body recombination toward these states \cite{Shotan14}. For example, three-body Efimov resonances have been experimentally observed through the enhancement of losses \cite{Kraemer06, Barontini09, Zacanti09}. In optical lattices, the engineering of three-body interactions was proposed via strong three-body losses and quantum Zeno effect \cite{Daley09, Mark20} or via a coherent coupling between two spin states \cite{Petrov14lattice}. 

Alternatively, an effective three-body interaction can be induced through a density dependant two-body coupling strength. This method requires an additional degree of freedom, that rapidly adjusts to the local density, and can be adiabatically eliminated. For a condensate in quasi-one-dimensional (1D) or quasi-2D geometries, the wave-function in the confined direction provides this additional degree of freedom \cite{Muryshev02, Mazets08, Merloti13}; its size slightly increases (decreases) for repulsive (attractive) two-body interactions. This effect leads to an effective attractive three-body coupling constant $g_3\propto -a^2$ in the equation of state in the reduced geometry. However, it remains a perturbative expansion, valid only when the three-body energy is a small correction 
as compared the two-body 
energy. Manifestations of these three-body interactions were nevertheless observed in a frequency shift of breathing oscillations 
in a quasi-2D geometry \cite{Merloti13} and in breaking integrability in quasi-1D gases \cite{Mazets08}. 

In this letter, we demonstrate that the additional spinor degree of freedom in coherently driven two-component condensate can similarly induce an effective three-body interaction after its adiabatic elimination. The method crucially relies on two facts: firstly, the scattering lengths in the dressed states depends on their spin composition \cite{Search01, Sanz21}; secondly, the spin composition is affected by density-induced mean-field shifts of the driven transition \cite{footnoteZibold, Zibold2010}. In contrast to condensates in reduced dimension, the two parameters in driven two-component condensates (the detuning frequency $\delta/2\pi$ and the Rabi-coupling frequency $\Omega/2\pi$) allow the independent control of the two-body and three-body coupling constants. The two-body interactions can thus be reduced such that the three-body interactions prevail in the equation of states. In addition, these three-body interactions appear at the mean-field level and can be made significantly larger in magnitude than the recently studied beyond-mean field three-body effects \cite{Petrov14, Lavoine21}.

Experimentally, we study two consequences of the effective attractive three-body interactions that appear in the lowest energy dressed-state of a driven two component $^{39}$K condensate. First, the radial breathing mode frequency of an elongated condensate, which is usually independent of the two-body interaction, exhibits a down shift. Second, we measure the threshold for radial collapse as a function of $\delta/\Omega$ and $\Omega$. We find that the condensate collapses despite a positive scattering length due to the attractive three-body interactions. A quantitative fit of experimental data requires taking into account the saturation of the interaction energy at high density, an effect beyond the three-body approximation. Importantly, we detect no increase of losses associated with the tuning of the two-body and three-body interactions in our system. 

\begin{figure}[h!]
\includegraphics[width=\columnwidth]{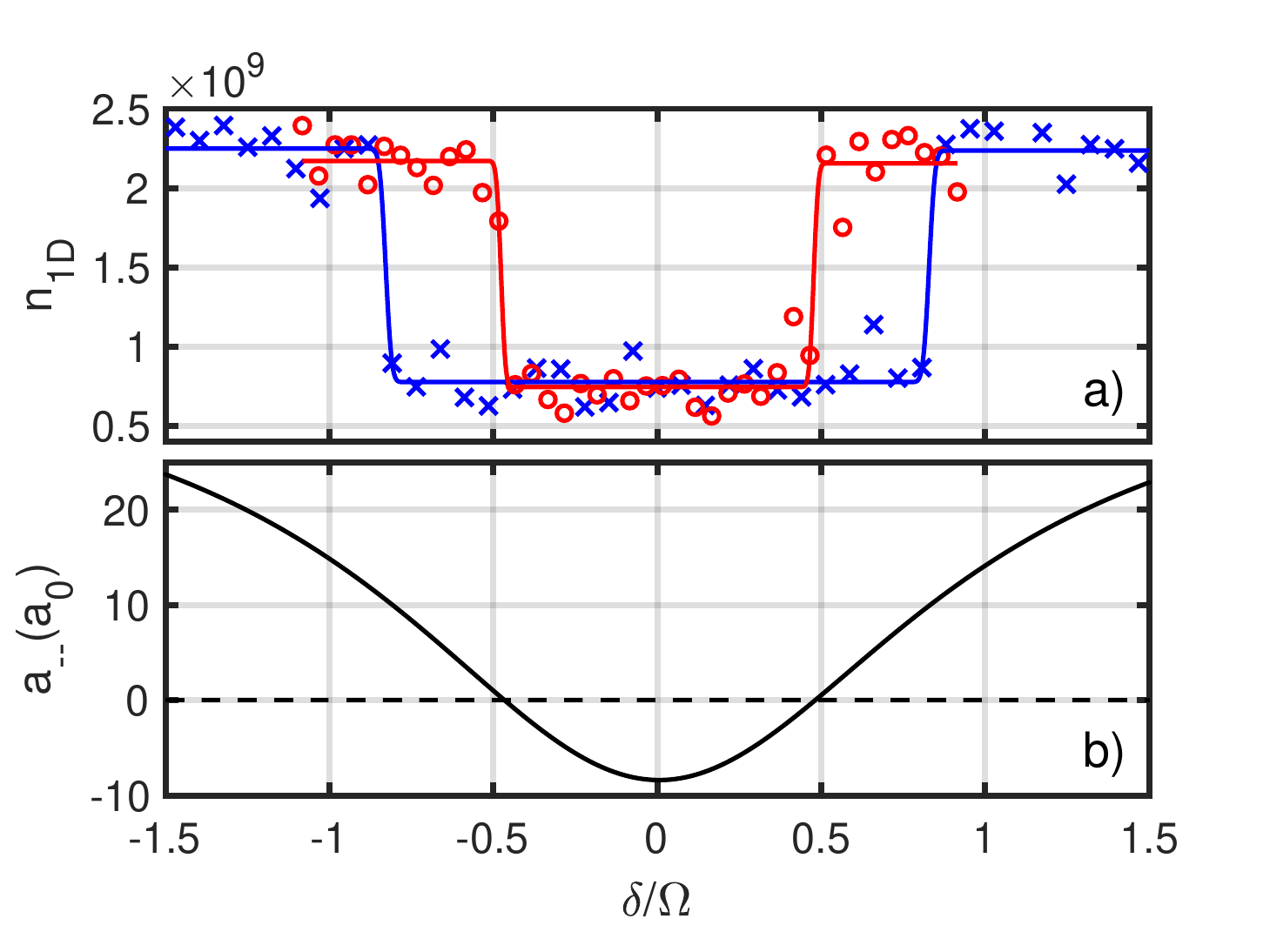}
\caption{(a) Remaining 1D density as a function of the final detuning $\delta/\Omega$. \textcolor{blue}{\boldmath$\times$} : $\Omega/2\pi=7.6$\,kHz, \textcolor{red}{\boldmath$\circ$} : $\Omega/2\pi=29.8$\,kHz. The curves are fits with the function $n_\textrm{coll}+(n_\textrm{1D}-n_\textrm{coll})\mathrm{erf}(\frac{\delta-\delta_c}{W \Omega})$, where $n_\textrm{coll}$, $n_\textrm{1D}$, $W$, and $\delta_c$ are free parameters. (b) Scattering length $a_{--}$ as a function of $\delta/\Omega$. \label{methodcollapse}}
\end{figure}

Let us consider a Bose gas in a volume $V$ formed by $N$ atoms of mass $m$ with two coupled internal states, $\sigma = \1,\2$. For simplicity, we suppose that interactions are symmetric, $g_{\1\1}=g_{\2\2}=g$, and we define  $\overline{g}=(g_{\1\1}-g_{\1\2})/2$,
with $g_{\sigma\sigma'}=4\pi\hbar^2a_{\sigma\sigma'}/m$ and $\hbar$ the reduced Planck constant. In an homogeneous system with density $n$, the mean-field energy for a condensate in the spinor state $(\phi_\1, \phi_\2)$ reads
\begin{gather}
\begin{split}
\nonumber
\frac{E_{MF}}{V}=-\frac{\hbar\Omega}{2}(\phi_\1^*\phi_\2+\phi_\2^*\phi_\1)+\frac{\hbar\delta}{2}(\left|\phi_\1\right|^2-\left|\phi_\2\right|^2)\\
+\sum_{\sigma\sigma'}\frac{g_{\sigma\sigma'}}{2}\left|\phi_\sigma\right|^2 \left|\phi_{\sigma'}\right|^2\textrm{.}
\end{split}
\end{gather} 
The ground state is found upon minimisation of the energy. The first term fixes the relative phase of the spinor $(\phi_\1, \phi_\2)=\sqrt{n}(\sin(\theta/2), \cos(\theta/2))$. The energy thus writes 
\begin{gather}
\frac{E_{MF}}{N}=-\frac{\hbar\Omega}{2} \sin(\theta)-\frac{\hbar\delta}{2}\cos(\theta)+\frac{gn}{2}-\frac{\overline{g}n}{2}\sin^2(\theta) \label{eqEMF}
\end{gather}
with $\theta \in [0, \pi]$ found upon minimisation. 
Up to first order in the ratio $\gamma = \frac{\overline{g}n}{\hbar\Omega}$, which compares the differential mean-field shift to the Rabi frequency, we find
\begin{equation}\label{theta}
\theta \approx \theta_0-2\gamma \frac{\delta/\Omega}{\left(1+\delta^2/\Omega^2\right)^{3/2}} \textrm { with }\textrm{cotan}(\theta_0)=\frac{\delta}{\Omega} \textrm{ ,}
\end{equation}
where $\gamma$ appears as the key parameter controlling the modification of the spinor degree of freedom. It results in the following mean-field energy (up to second order in $\gamma$)
\begin{gather}
\frac{E_{MF}}{N}\approx -\frac{\hbar\sqrt{\Omega^2+\delta^2}}{2}+g_2\frac{n}{2}+g_3\frac{n^2}{3}, \label{energy}\\
\text{with } g_2=g-\frac{\overline{g}}{1+\delta^2/\Omega^2}\\
\text{and } g_3=-\frac{3\overline{g}^2}{\hbar\Omega}\frac{\delta^2/\Omega^2}{\left(1+\delta^2/\Omega^2\right)^{5/2}}.
\end{gather}
The first term in Eq.\,\ref{energy} corresponds to the energy of the lower dressed-state $|-\rangle$ in the absence of interaction. $g_2=4\pi\hbar^2a_{--}/m$ corresponds to the two-body coupling constant in the unperturbed $|-\rangle$ state, i.e. with $\theta=\theta_0$. It is  solely determined by the ratio $\delta/\Omega$ (see Fig.\,\ref{methodcollapse}b). $g_3$ is an attractive three-body coupling constant appearing because of the mean-field-induced change in $\theta$ (see Eq.\,\ref{theta}). 
 It is zero both for large absolute value of $\delta/\Omega$ when the two states are uncoupled but also for $\delta=0$ as in an equal mixture the energy is already extremal at $\theta=\theta_0$. Interestingly, $g_3$ can be independently controlled from $g_2$ through the value of $\Omega$. However $\Omega$ can not be made arbitrarily small as the adiabatic following of the dressed state requires $\dot{\gamma} \ll \Omega$.


Note that the energy expansion in powers of the density (Eq.\,\ref{energy}) is only valid for $\gamma\ll 1$. In the opposite limit $\gamma\gg 1$, the interaction energy $\frac{gn}{2}-\frac{\overline{g}n}{2}\sin^2(\theta)$ in Eq.\,\ref{eqEMF} saturates for $\theta=\pi/2$ and $g_\textrm{min}=(g-\overline{g})$. For $\gamma\sim 1$, the interaction energy cannot be reduced to a sum of two and three body coupling terms. In this case, the energy $E_{MF}$ needs to be numerically calculated by minimizing equation \ref{eqEMF} as a function of $\theta$.

\begin{figure}[h!]
\includegraphics[width=\columnwidth]{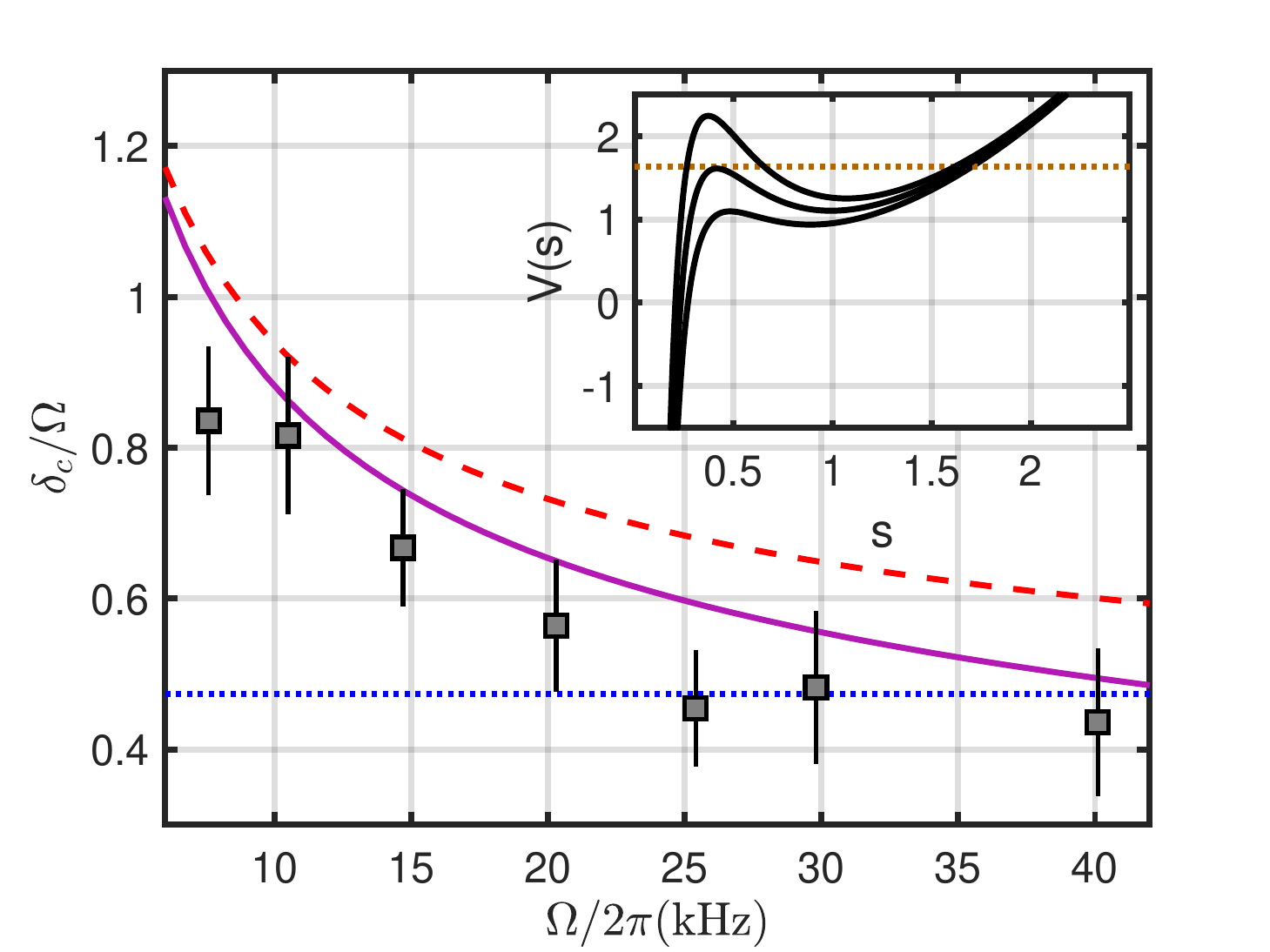}
\caption{Collapse threshold as a function of the Rabi frequency $\Omega/2\pi$. The squares are the experimental data. The dotted blue (green) line at $\delta/\Omega=0.47$ corresponds to $a_{--}=0$. The solid purple (dashed red) line corresponds to the theoretical prediction taking into account the mean-field effect on the internal state with (and without) the renormalization of the two-body interaction (see text). Inset: Effective potentials $V$ for $\Omega/2\pi=30\,$kHz for $\delta/\Omega=0.54$ (top curve), $\delta/\Omega=0.49=\delta_c/\Omega$ (middle curve), $\delta/\Omega=0.44$ (bottom curve). The dotted line is the initial energy for the middle curve corresponding to $s(0)=1.7$. \label{threshold_results}}
\end{figure}

We now turn to the experimental observation of the three-body interactions. We work with the second and third lowest Zeeman states of the lowest manifold of $^{39}\rm{K}$, namely  $\ket{\1}=\ket{\rm{F}=1,\rm{m_{F}}=-1}$ and $\ket{\2}=\ket{\rm{F}=1,\rm{m_{F}}=0}$. At a magnetic field of $54.690(1)\,\rm{G}$, the three relevant scattering lengths are $a_{\1\1}=37.9 a_0 \approx a_{\2\2}=36.9,a_0$ \cite{unbalanced}, and  $a_{\1\2}=-54.2\,a_0$, where $a_0$ is the atomic Bohr radius \cite{Tiemann20}. With these specific parameters, the scattering length $a_{--}$ in the $|-\rangle$ dressed-state  can be tuned down to -8.4\,$a_0$ for $\delta\approx 0$ and has zero crossings at $|\delta/\Omega| \approx 0.47$ (see Fig.\,\ref{threshold_results}). Because of our rms magnetic field noise of 0.8(2)\,mG corresponding to 0.56(14)\,kHz, we choose to work with $\Omega/2\pi \geq 7$\,kHz in order to keep a good control of the parameter $\delta/\Omega$. For this value of $\Omega$ and $\delta/\Omega\approx 0.8$, the minimum three-body coupling constant is $g_3/\hbar=-3\times 10^{-38}$\,m$^6$.s$^{-1}$. This value is typically larger by a factor $\sim$100 as compared to the dominant three-body loss coefficient $K_3^{\2\2\2}/6\approx 3\times 10^{-40}$\,m$^6$.s$^{-1}$ in our potassium mixtures \cite{Cheiney18}. The hypervolume $D$ for our parameters is thus essentially real with $\Re{(D)} \gg \Im{(D)}$. 

 The experiment starts with a quasi pure BEC with $\sim 1.4\times 10^5$ atoms in state $\ket{\1}$ in a cigar shaped trap with frequencies $(\omega_\perp,\omega_\parallel)/2\pi=(300,16.4)\; \rm{Hz}$. The condensate in the dressed state $\ket{-}$ is then  prepared in an adiabatic passage, in which the radio-frequency (rf) detuning is swept from $7.5\,\Omega$ to its final value $\delta$. Its shape and duration of 0.4\,ms are chosen in order to be adiabatic with respect to the internal-state dynamics but it is short as compare to $2\pi/\omega_\perp$. As a consequence, the rf sweep is equivalent to a quench of the interaction parameters and it induces some dynamics of the cloud.  In the longitudinal direction, the evolution is slow and we neglect it on the 15\,ms time scale of our experiment \cite{longi}. We focus our analysis on the radial dynamics of the condensate in its central part where the 1D density $n_\textrm{1D}$ is approximately constant.

\begin{figure}[h]
\includegraphics[width=\columnwidth]{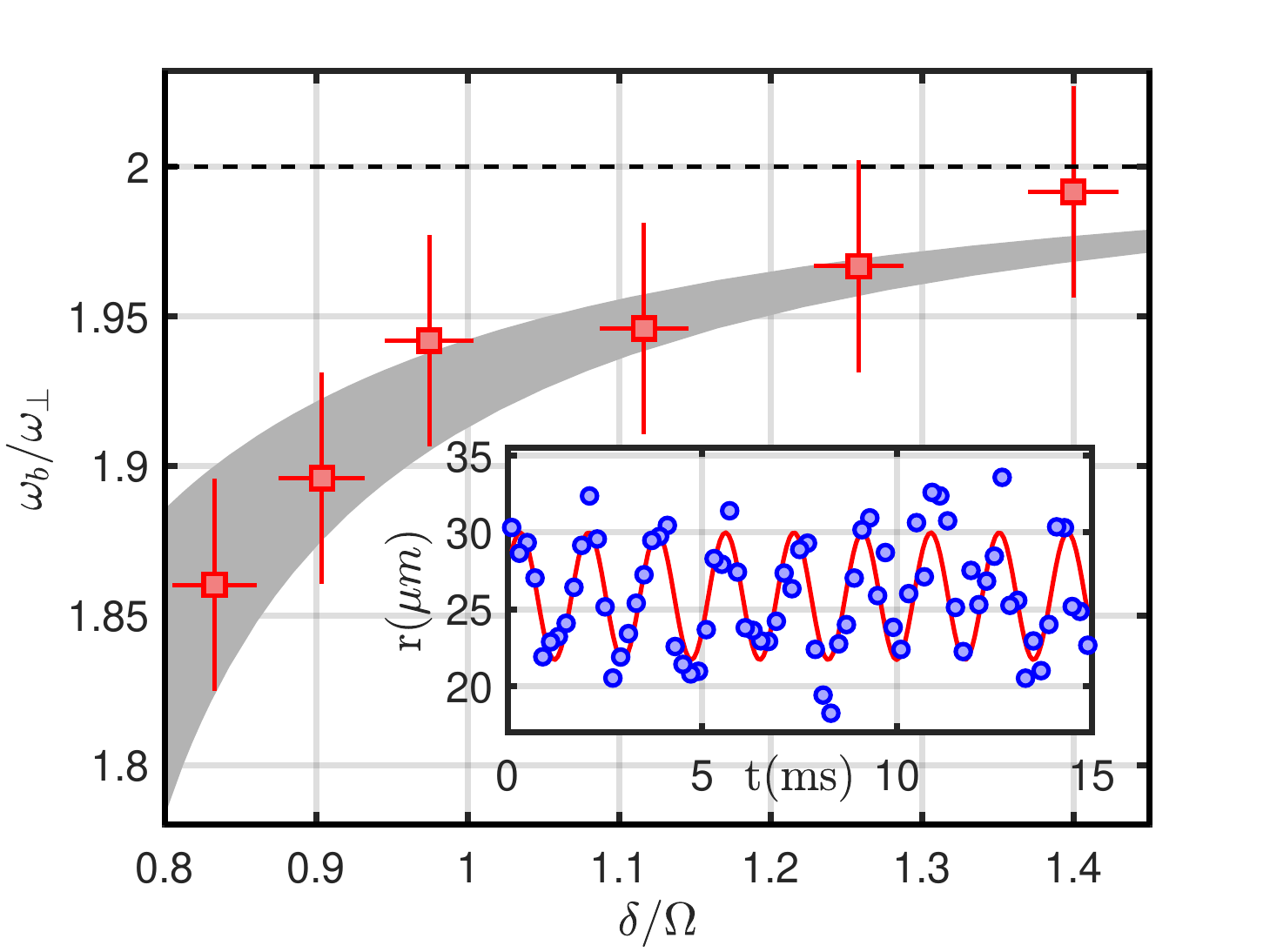}
\caption{Breathing frequency at Rabi frequency $\Omega = 25.4\,kHz$ as a function of the  detuning $\delta/\Omega$. The points correspond to the experimental data. The vertical error bars correspond to a 1.5\% uncertainty in the measured frequency. The horizontal error bars are linked to our 0.8(2)\,mG magnetic field fluctuations. The shaded area corresponds to the theoretical estimates for $2.3\times 10^9$\,m$^{-1}<n_\textrm{1D}<2.65\times 10^9\,$m$^{-1}$ taking into account the uncertainty in the value of $n_\textrm{1D}$ due to experimental fluctuations, losses, and uncertainty in the detection efficiency. Inset: radial breathing oscillations for $\delta/\Omega=0.9$. The rms radius $r$ of the gas is measured as a function of the wait time $t$ after 9.7\,ms of free expansion including an initial 0.4\,ms rf sweep back to $\delta=7.5\Omega$. 
\label{breathing}}
\end{figure}

In a first series of experiments, we chose parameters ($\Omega/2\pi= 25.4\,$kHz and $\delta/\Omega>0.8$) for which, we observe small amplitude breathing oscillation of the radial size (see inset in Fig.\ref{breathing}) \cite{small_osc}. 
On the 15\,ms time scale of the experiment, we find that the atom number is reduced by a maximum of 20\%. Interestingly, we measure a reduction of the breathing mode frequency as $\delta/\Omega$ is decreased from 1.4 to 0.8 (see Fig.\,\ref{breathing}), whereas, in the absence of three-body interaction, it is expected to be constant and equal to $2 \omega_\perp$ independently of the two-body contact interaction due to a hidden symmetry of the Hamiltonian under scale transformation \cite{Kagan96, Pitaevskii97, Chevy02}. Our observation of a frequency downshift thus points toward a role of the three-body interactions.  

Let us now compare the measured frequencies to theoretical expectations for a condensate for which the equation of state is given by Eq.\,\ref{energy}. 
In a variational and scaling approach \cite{Pethick, Jorgensen18}, the frequency of small breathing oscillations writes 
\be
\omega_\textrm{b}=2\omega_\perp\sqrt{1+E_3/E_\textrm{pot}},
\ee
where $E_3<0$ and $E_\textrm{pot}$ are the three-body and potential energy in the equilibrium state. We calculate these two quantities from imaginary time evolution of a 2D Gross-Pitaevskii equation \cite{GPELab} and deduce the value of the breathing mode frequency (see Fig.\,\ref{breathing}).  Within the experimental error bars, we find a perfect agreement with the measured values and we thus attribute the breathing mode frequency downshift to the attractive three-body interactions. Note that in the explored range,  decreasing $\delta/\Omega$ corresponds both to a decrease of $g_2$ and to an increase of $|g_3|$.

By lowering further the value of $\delta/\Omega$, we observe large losses that rapidly occur around $\sim$1\,ms after the beginning of the RF sweep, i.e. when the condensate has shrunk to a high density. In order to study this behavior, we wait 3\,ms after the sweep and plot the remaining central 1D density as a function of $\delta/\Omega$ for two values of $\Omega$ (see Fig.\,\ref{methodcollapse}a). At large value of $|\delta/\Omega|$, there are few losses and the 1D density is close to the initial one. On the contrary, for low value of $|\delta/\Omega|$, the 1D density is observed to be reduced by a  factor $\sim$3. Interestingly, the losses appear sharply as a function of $|\delta/\Omega|$ and we interpret this behavior as originating from a radial collapse of the cloud, which is certainly expected for $\delta/\Omega\approx 0$ where the minimum scattering length is $a_{--}=-8.4\,a_0<0$. In the following, we do not try to precisely understand the collapse dynamics, including the role of losses, but rather focus our analysis on the threshold values $\delta_c/\Omega$ below which a collapse occurs. 

The collapse thresholds $\delta_c/\Omega$ are plotted as a function of $\Omega$ in Fig.\, \ref{threshold_results} and are found to be larger for lower values of $\Omega$. Such a behavior reveals the role of three-body interactions in the radial collapse as $g_2$ solely depends on the ratio $\delta/\Omega$. Moreover, for $\Omega/2\pi<20\,$kHz the collapse is observed for $\delta/\Omega > 0.47$, which corresponds to a positive scattering length $a_{--}$, $i.e.$ repulsive two-body interactions (see Fig.\,\ref{methodcollapse}b). As an example, for $\Omega=7.6\,$kHz, $\delta_c/\Omega=0.82(5)$ corresponds to $a_{--}\approx$10\,$a_0$ (see Fig.\,\ref{methodcollapse}).

In order to quantitatively interpret our findings,  we develop a model which assumes, for simplicity, a Gaussian radial density profile at all times $t$ with a rms radial size $s(t) a_\textrm{ho}/\sqrt{2}$, where $a_\textrm{ho}=\sqrt{\hbar/m\omega_\perp}$. The energy of the system can then be cast as  
\be
\label{Energy}
\frac{E_{MF}}{N\hbar\omega_r}=\frac{\dot{s}^2}{2}+\underbrace{\frac{s^2}{2}+\frac{1}{2s^2}+\frac{n_{1D}a_{--}}{s^2}+V_{\textrm{corr}}(s)}_{V(s)}, 
\ee
where the terms on the right hand side respectively corresponds to current kinetic energy, potential energy, zero-point kinetic energy \cite{thetagrad}, two-body interactions and corrections to the interaction energy due to the mean-field induced change in the internal state. For $\gamma \ll 1$, this last term corresponds to attractive three-body interactions $V_{\textrm{corr}}(s)\propto g_3n_{1D}^2/s^4$. Since we only see important losses when the collapse has occurred, we do not include a loss term in the initial dynamics. In this framework, the latter reduces to the one of a classical particle in an effective potential $V(s)$ given by the four last terms in Eq.\,\ref{Energy}. 

Typical effective potentials $V(s)$ close to the collapse threshold are plotted in the inset of figure \ref{threshold_results}. They exhibit a local maximum for low value of $s$ that may be overcome or not depending on the initial energy, which is given by the initial rms size of the cloud. The latter is numerically computed \cite{GPELab} and is 1.1\,$\mu$m corresponding to $s(0)=1.7$.
 Note that at the position of the maximum of $V(s)$, the density is such that $\gamma \gtrsim 1$ and we have to rely on numerical calculation of $V_{\textrm{corr}}(s)$ for a precise determination of the effective potential $V(s)$. For each value of $\Omega$, we find the collapse threshold $\delta_c/\Omega$ for which the local maximum of $V(s)$ is equal to $V(s(0))$. This model for the collapse threshold (dashed curve in Fig.\,\ref{threshold_results}) captures the trend of the threshold values $\delta_c/\Omega$ but slightly overestimates them. As an improvement to our model, we may introduce repulsive beyond-mean field effects in the equation of state that are neglected in our mean-field approach \cite{Lavoine21}. Since they are complex to evaluate in full generality, as a first estimate, we solely take into account the additional two-body correction associated with the modification of $a_{--}$ originating from virtual transitions to the high-energy dressed state $|+\rangle$ \cite{Lavoine21, footnoteLHY}. A better match to the experimental data is then obtained (see the solid line in Fig.\,\ref{threshold_results}).

To conclude, we have shown that a Rabi-coupled two component Bose-Einstein condensate with different scattering lengths offers a way to induce an attractive three-body term in the equation of state. The latter appears, at the mean-field level, because of a density-dependent detuning of the drive and is tunable through the strength of the Rabi drive. The attractive three-body energy can be made large as compared to the three-body loss rate which does not depend on $\Omega$. Experimentally, we study two striking consequences of the attractive three-body term: a shift of the radial breathing mode frequency and the observation of radial collapses despite repulsive two-body couplings $g_2>0$. The precise collapse dynamics including spin dynamics, losses and beyond-mean-field effects could be the object of future studies.

Finally, we discuss the possibility to create a repulsive three-body term with a condensate in the excited dressed state $|+\rangle$. In this case, $\theta \in [-\pi, 0]$ is found upon maximization of equation \ref{eqEMF} and the sign of $g_3$ is reversed. Unfortunately, a condensate in $|+\rangle$ suffers from large two-body losses \cite{Sanz21}. The two-body loss rate is $\Gamma \propto n \overline{a}^2 \sqrt{\Omega}\propto  n \overline{a}^2/l_\Omega$, where $\overline{a}=a_{\1\1}-a_{\1\2}$ and $l_\Omega=\sqrt{\hbar/m\Omega}$ is a length scale associated with $\Omega$. Reducing the value of $\Omega$ would open a window where the repulsive three-body energy could dominate over the two-body loss rate for $E_3/\hbar\Gamma \propto nl_\Omega^3\gg 1 \gtrsim n \overline{a} \l_\Omega^2$, where the second inequality ensures that $\gamma \lesssim 1$. Experimentally, this   requires a reduction of the role of magnetic fluctuations. 
Repulsive three-body interactions produced in this manner would offer an alternative way to create gaseous (potentially more strongly bound) droplets as compared to beyond-mean field effects \cite{Petrov2015, PfauDroplet, FerlainoDroplet, Cabrera2018, Semeghini2018, Guo2021}. Quantum droplets \cite{Sekino18} and few-body bound states \cite{Nishida18, Guijarro18} were also recently discussed in the case of 1D bosons with three-body interactions. 

\begin{acknowledgements}
We thank D. Cl\'ement, D. Petrov, S. Kokkelmans, L. Tarruell, C. Westbrook, and W. Zwerger for useful discussions. This  research  has  been  supported  by  CNRS,  Minist\`ere  de  l'Enseignement  Sup\'erieur  et  de  la  Recherche, Labex PALM, Region Ile-de-France  in  the  framework  of  Domaine d'Intérêt Majeur  Sirteq, Paris-Saclay in the framework of IQUPS, ANR Droplets (19-CE30-0003), Simons foundation (award number 563916:  localization of waves).
\end{acknowledgements}


\begin{thebibliography}{0}%
\makeatletter
\providecommand \@ifxundefined [1]{%
 \@ifx{#1\undefined}
}%
\providecommand \@ifnum [1]{%
 \ifnum #1\expandafter \@firstoftwo
 \else \expandafter \@secondoftwo
 \fi
}%
\providecommand \@ifx [1]{%
 \ifx #1\expandafter \@firstoftwo
 \else \expandafter \@secondoftwo
 \fi
}%
\providecommand \natexlab [1]{#1}%
\providecommand \enquote  [1]{``#1''}%
\providecommand \bibnamefont  [1]{#1}%
\providecommand \bibfnamefont [1]{#1}%
\providecommand \citenamefont [1]{#1}%
\providecommand \href@noop [0]{\@secondoftwo}%
\providecommand \href [0]{\begingroup \@sanitize@url \@href}%
\providecommand \@href[1]{\@@startlink{#1}\@@href}%
\providecommand \@@href[1]{\endgroup#1\@@endlink}%
\providecommand \@sanitize@url [0]{\catcode `\\12\catcode `\$12\catcode
  `\&12\catcode `\#12\catcode `\^12\catcode `\_12\catcode `\%12\relax}%
\providecommand \@@startlink[1]{}%
\providecommand \@@endlink[0]{}%
\providecommand \url  [0]{\begingroup\@sanitize@url \@url }%
\providecommand \@url [1]{\endgroup\@href {#1}{\urlprefix }}%
\providecommand \urlprefix  [0]{URL }%
\providecommand \Eprint [0]{\href }%
\providecommand \doibase [0]{http://dx.doi.org/}%
\providecommand \selectlanguage [0]{\@gobble}%
\providecommand \bibinfo  [0]{\@secondoftwo}%
\providecommand \bibfield  [0]{\@secondoftwo}%
\providecommand \translation [1]{[#1]}%
\providecommand \BibitemOpen [0]{}%
\providecommand \bibitemStop [0]{}%
\providecommand \bibitemNoStop [0]{.\EOS\space}%
\providecommand \EOS [0]{\spacefactor3000\relax}%
\providecommand \BibitemShut  [1]{\csname bibitem#1\endcsname}%
\let\auto@bib@innerbib\@empty
\end{thebibliography}%


\begin{thebibliography}{0}
\bibitem{Pitaevskii03}
L.Pitaevskii and S. Stringari, {\it Bose-Einstein Condensation and superfluidity}, Oxford University Press (2003).

\bibitem{Chin10}
C. Chin, R. Grimm, P. Julienne, and E. Tiesinga, Feshbach resonances in ultracold gases, Rev. Mod. Phys. {\bf 82}, 1225 (2010).

\bibitem{Bloch08}
I. Bloch, J. Dalibard, and W. Zwerger, Many-body physics with ultracold gases,
Rev. Mod. Phys. {\bf 80}, 885 (2008).

\bibitem{Greiner02}
M. Greiner, O. Mandel, T. Esslinger, T.W. H\"ansch and I. Bloch, Quantum phase transition from a superfluid to a Mott insulator in a gas of ultracold atoms, Nature {\bf 415}, 39 (2002).

\bibitem{Regal04}
C.A. Regal, M. Greiner, and D.S. Jin,
Observation of Resonance Condensation of Fermionic Atom Pairs, Phys. Rev. Lett. {\bf 92}, 040403 (2004).

\bibitem{Zwierlein04}
M.W. Zwierlein, C.A. Stan, C.H. Schunck, S.M.F. Raupach, A. J. Kerman, and W. Ketterle, 
Condensation of Pairs of Fermionic Atoms near a Feshbach Resonance, Phys. Rev. Lett. {\bf 92}, 120403 (2004).

\bibitem{Bartenstein04}
M. Bartenstein, A. Altmeyer, S. Riedl, S. Jochim, C. Chin, J. Hecker Denschlag, and R. Grimm, 
Crossover from a Molecular Bose-Einstein Condensate to a Degenerate Fermi Gas, Phys. Rev. Lett. {\bf 92}, 120401 (2004).

\bibitem{Bourdel04}
T. Bourdel, L. Khaykovich, J. Cubizolles, J. Zhang, F. Chevy, M. Teichmann, L. Tarruell, S.J.J.M.F. Kokkelmans, and C. Salomon, 
Experimental Study of the BEC-BCS Crossover Region in Lithium 6,
Phys. Rev. Lett. {\bf 93}, 050401 (2004).

\bibitem{Zwerger16}
W. Zwerger, Strongly Interacting Fermi Gases, Proceedings of the International School of Physics "Enrico Fermi" edited by M. Inguscio, W. Ketterle, S. Stringari and G. Roati (IOS Press, Amsterdam; SIF Bologna) 2016, pp. 63.

\bibitem{Wu59}
T.T. Wu, Ground State of a Bose System of Hard Spheres, Phys. Rev. {\bf 115}, 1390 (1959).

\bibitem{Kohler02}
T. K\"ohler, Three-body problem in a dilute Bose-Einstein condensate, Phys. Rev. Lett. {\bf 89}, 210404 (2002).

\bibitem{Hammer13}
H.W. Hammer, A. Nogga, and A. Schwenk, Colloquium: Three-body forces: From cold atoms to nuclei, Rev. Mod. Phys. {\bf 85}, 197 (2013).



\bibitem{Gammal00}
A. Gammal, T. Frederico, L. Tomio, and P. Chomaz, Atomic Bose-Einstein condensation with three-body interactions and collective excitations, Journal of Physics B: Atomic, Molecular and Optical Physics {\bf 33}, 4053 (2000).

\bibitem{Gammal00b}
A. Gammal, T. Frederico, L. Tomio, and F.K. Abdullaev,
Stability analysis of the D-dimensional nonlinear Schrödinger equation with trap and two- and three-body interactions, Phys. Lett. A {\bf 267}, 305 (2000).

\bibitem{Kumar10}
V.R. Kumar, R. Radha, and M. Wadati, Phase Engineering and Solitons of Bose–Einstein Condensates with Two- and Three-Body Interactions, 
J. Phys. Soc. Jpn. {\bf 79}, 074005 (2010).

\bibitem{Crosta12}
M. Crosta, S. Trillo, and A. Fratalocchi,
The Whitham approach to dispersive shocks in systems with cubic-quintic nonlinearities, New J. Phys. {\bf 14}, 093019 (2012).

\bibitem{Al-Jibbouri13}
H. Al-Jibbouri, I. Vidanović, A. Balaž, and A. Pelster, Geometric resonances in Bose–Einstein condensates with two- and three-body interactions, J. Phys. B: At. Mol. Opt. Phys. {\bf 46}, 065303 (2013).

\bibitem{Killip17}
R. Killip, T. Oh, O. Pocovnicu, and M. Visan, Solitons and Scattering for the Cubic-Quintic Nonlinear Schrodinger Equation on R$^3$, Arch. Rational Mech. Anal. {\bf 225}, 469 (2017).


\bibitem{Zloshchastiev17}
K.G. Zloshchastiev, Stability and Metastability of Trapless Bose-Einstein Condensates and Quantum Liquids,
Zeitschrift f\"ur Naturforschung A {\bf 72}, 677 (2017). 

\bibitem{Bulgac02}
A. Bulgac, Dilute Quantum Droplets, Phys. Rev. Lett. {\bf 89}, 050402 (2002).


\bibitem{Tan08}
S. Tan, Three-boson problem at low energy and implications for dilute Bose-Einstein condensates, Phys. Rev. A {\bf 78}, 013636 (2008).

\bibitem{Zhu17}
S. Zhu and S. Tan, Three-body scattering hypervolumes of particles with short-range interactions, arXiv:1710.04147 (2017).

\bibitem{Mestrom19}
P.M.A. Mestrom, V.E. Colussi, T. Secker, and S.J.J.M.F. Kokkelmans, Scattering hypervolume for ultracold bosons from weak to strong interactions, 
Phys. Rev. A {\bf 100}, 050702(R) (2019).

\bibitem{Incao18}
J. P. D'Incao, Few-body physics in resonantly interacting ultracold quantum gases, J. Phys. B: At. Mol. Opt. Phys. {\bf 51},  043001 (2018).

\bibitem{Zwerger19}
W. Zwerger, Quantum-unbinding near a zero temperature liquid-gas transition,
J. Stat. Mech. Theory and Experiment {\bf 10}, 103104 (2019).

\bibitem{Shotan14}
Z. Shotan, O. Machtey, S. Kokkelmans, and L. Khaykovich, Three-Body Recombination at Vanishing Scattering Lengths in an Ultracold Bose Gas, Phys. Rev. Lett. {\bf 113}, 053202 (2014).

\bibitem{Kraemer06}
T. Kraemer, M. Mark, P. Waldburger, J. G. Danzl, C. Chin, B. Engeser, A. D. Lange, K. Pilch, A. Jaakkola, H.-C. N\"agerl, and  R. Grimm, Evidence for Efimov quantum states in an ultracold gas of caesium atoms, Nature {\bf 440}, 315 (2006).

\bibitem{Barontini09}
G. Barontini, C. Weber, F. Rabatti, J. Catani, G. Thalhammer, M. Inguscio, and F. Minardi, Observation of Heteronuclear Atomic Efimov Resonances, Phys. Rev. Lett. {\bf 103}, 043201 (2009).

\bibitem{Zacanti09}
M. Zaccanti, B. Deissler, C. D’Errico, M. Fattori, M. Jona-Lasinio, S. M\"uller, G. Roati, M. Inguscio, and G. Modugno, Observation of an Efimov spectrum in an atomic system, Nature Physics {\bf 5}, 586 (2009).

\bibitem{Daley09}
A. J. Daley, J. M. Taylor, S. Diehl, M. Baranov, and P. Zoller, Atomic Three-Body Loss as a Dynamical Three-Body Interaction,
Phys. Rev. Lett. {\bf 102}, 040402 (2009).

\bibitem{Mark20}
M. J. Mark, S. Flannigan, F. Meinert, J. P. D'Incao, A. J. Daley, and H.-C. N\"agerl, Interplay between coherent and dissipative dynamics of bosonic doublons in an optical lattice, Phys. Rev. Research {\bf 2}, 043050 (2020).

\bibitem{Petrov14lattice}
D. S. Petrov, Elastic multibody interactions on a lattice,
Phys. Rev. A {\bf 90}, 021601(R) (2014).


\bibitem{Muryshev02}
A. Muryshev, G. V. Shlyapnikov, W. Ertmer, K. Sengstock, and
M. Lewenstein, Dynamics of Dark Solitons in Elongated Bose-
Einstein Condensates, Phys. Rev. Lett. {\bf 89}, 110401 (2002).

\bibitem{Mazets08}
I. E. Mazets, T. Schumm, and J. Schmiedmayer, Breakdown of
Integrability in a Quasi-1D Ultracold Bosonic Gas, Phys. Rev.
Lett. {\bf 100}, 210403 (2008).

\bibitem{Merloti13}
K. Merloti, R. Dubessy, L. Longchambon, M. Olshanii, and H. Perrin, Breakdown of scale invariance in a quasi-two-dimensional Bose gas due to the presence of the third dimension, Phys. Rev. A {\bf 88}, 061603(R) (2013).

\bibitem{Search01}
C. P. Search and P. R. Berman, Manipulating the speed of sound in a two-component Bose-Einstein condensate, Phys. Rev. A {\bf 63}, 043612(2001).

\bibitem{Sanz21}
J. Sanz, A. Fr\"olian, C. S. Chisholm, C. R. Cabrera, L. Tarruell, Interaction control and bright solitons in coherently-coupled Bose-Einstein condensates, arXiv:1912.06041 (2019).


\bibitem{footnoteZibold}
 The same differential mean-field shift is at the origin of the blockade of internal Josephson oscillations \cite{Zibold2010}.

\bibitem{Zibold2010}
T. Zibold, E. Nicklas, C. Gross, and M. K. Oberthaler, Classical Bifurcation at the Transition from Rabi to Josephson Dynamics,
Phys. Rev. Lett. {\bf 105}, 204101 (2010).

\bibitem{Petrov14}
D. S. Petrov, Three-Body Interacting Bosons in Free Space, Phys. Rev. Lett. {\bf 112}, 103201 (2014).

\bibitem{Lavoine21}
L. Lavoine, A. Hammond, A. Recati, D. Petrov, and T. Bourdel, Beyond-mean-field effects in Rabi-coupled two-component Bose-Einstein condensate, Phys. Rev. Lett. {\bf 127}, 203402 (2021).



\bibitem{unbalanced}
We have checked that the slight unbalanced between $a_{\1\1}$ and $a_{\2\2}$ does not quantitatively play a role in our experimental findings and we simply consider the average value in the analysis.  

\bibitem{Tiemann20}
E. Tiemann, P. Gersema, K.K. Voges, T. Hartmann, A. Zenesini, and S. Ospelkaus, Beyond Born-Oppenheimer approximation in ultracold atomic collisions,
Phys. Rev. Research {\bf 2}, 013366 (2020).

\bibitem{Cheiney18}
P. Cheiney, C. R. Cabrera, J. Sanz, B. Naylor, L. Tanzi, and L. Tarruell, 
Bright Soliton to Quantum Droplet Transition in a Mixture of Bose-Einstein Condensates, 
Phys. Rev. Lett.  {\bf 120}, 135301 (2018).


\bibitem{longi}
Actually, we adjust the longitudinal trapping frequency after the rf sweep in order to detect no longitudinal dynamics. 

\bibitem{small_osc}
We have reduced the oscillation amplitude by applying a longer rf sweep and found no change in the breathing mode frequency. 

\bibitem{Kagan96}
Y. Kagan, E.L. Surkov, and G.V. Shlyapnikov, Evolution of a Bose-condensed gas under variations of the confining potential, Phys.Rev. A {\bf 54}, R1753 (1996).

\bibitem{Pitaevskii97}
L.P. Pitaevskii and A. Rosch, Breathing modes and hidden symmetry of trapped atoms in two dimensions, Phys. Rev. A {\bf 55}, R853 (1997).

\bibitem{Chevy02}
F. Chevy, V. Bretin, P. Rosenbusch, K. W. Madison, and J. Dalibard, Transverse Breathing Mode of an Elongated Bose-Einstein Condensate, 
Phys. Rev. Lett. {\bf 88}, 250402 (2002).

\bibitem{Pethick}
C. Pethick and H. Smith,  {\it Bose–Einstein Condensation in Dilute Gases},  Cambridge University Press (2008).


\bibitem{Jorgensen18}
N.B. Jørgensen, G.M. Bruun, and J.J. Arlt, Dilute Fluid Governed by Quantum Fluctuations,
Phys. Rev. Lett. {\bf 121}, 173403 (2018).

\bibitem{GPELab}
X. Antoine and R. Duboscq, GPELab, a Matlab Toolbox to Solve Gross-Pitaevskii Equations I:computation of stationary solutions, Computer Physics Communications {\bf 185}, 2969 (2014).

\bibitem{thetagrad}
We have omitted in the description the energy associated to the gradient of $\theta$ \cite{Matthews99} which adds a negligible kinetic energy in our experimental configuration.



\bibitem{footnoteLHY}
Strictly, higher-order beyond-mean field corrections will make $V(s)$ to increase at low $s$ and prevent the collapse. However, this would occur at such high densities that losses will anyway occur before reaching such a small size, leaving our conclusion unchanged. 



\bibitem{Petrov2015}
D.S. Petrov, 
Quantum mechanical stabilization of a collapsing Bose-Bose mixture,
Phys. Rev. Lett. {\bf 115}, 155302 (2015).

\bibitem{PfauDroplet}
I. Ferrier-Barbut, H. Kadau, M. Schmitt, M. Wenzel, and T. Pfau, Observation of Quantum Droplets in a Strongly Dipolar Bose Gas, Phys. Rev. Lett.  {\bf 116}, 215301 (2016).

\bibitem{FerlainoDroplet}
L. Chomaz, S. Baier, D. Petter, M. J. Mark,
F. W\"achtler, L. Santos, and F. Ferlaino, Quantum-Fluctuation-Driven Crossover from a Dilute Bose-Einstein Condensate to a Macrodroplet in a Dipolar Quantum Fluid, Phys. Rev. X
{\bf 6}, 041039 (2016).


\bibitem{Cabrera2018}
R. Cabrera, L. Tanzi, J. Sanz, B. Naylor, P. Thomas, P. Cheiney, and L. Tarruell,
Quantum liquid droplets in a mixture of Bose-Einstein condensates, 
Science  {\bf 359}, 301 (2018).

\bibitem{Semeghini2018}
G. Semeghini, G. Ferioli, L. Masi, C. Mazzinghi, L. Wolswijk, F. Minardi, M. Modugno, G. Modugno, M. Inguscio, and M. Fattori, Self-Bound Quantum Droplets of Atomic Mixtures in Free Space, 
Phys. Rev. Lett.  {\bf 120}, 235301 (2018).

\bibitem{Guo2021}
Z. Guo, F. Jia, L. Li, Yi. Ma, J.M. Hutson, X. Cui, and D. Wang, Lee-Huang-Yang effects in the ultracold mixture of $^{23}$Na and $^{87}$Rb with attractive interspecies interactions, arXiv:2105.01277.

\bibitem{Sekino18}
Y. Sekino and Y. Nishida, Quantum droplet of one-dimensional bosons with a three-body attraction, Phys. Rev. A {\bf 97}, 011602(R) (2018).

\bibitem{Nishida18}
Y. Nishida, Universal bound states of one-dimensional bosons with two- and three-body attractions, Phys. Rev. A {\bf 97}, 061603(R) (2018).

\bibitem{Guijarro18}
G. Guijarro, A. Pricoupenko, G.E. Astrakharchik, J. Boronat, and D.S. Petrov, One-dimensional three-boson problem with two- and three-body interactions, Phys. Rev. A {\bf 97}, 061605(R) (2018).

\bibitem{Matthews99}
M.R. Matthews, B.P. Anderson, P.C. Haljan, D.S. Hall, M.J. Holland, J.E. Williams, C.E. Wieman, and E.A. Cornell, Watching a Superfluid Untwist Itself: Recurrence of Rabi Oscillations in a Bose-Einstein Condensate, Phys. Rev. Lett. {\bf 83}, 3358 (1999).
\end{thebibliography}
\end{document}